\documentclass[useAMS,usenatbib]{mn2e}
\usepackage{times}
\usepackage{epsfig}
\usepackage{amssymb}
\usepackage{amsmath} 
\usepackage{aas_macros}
\title[Dust formation in the remnant of SN\,1987A]
{The timing and location of dust formation in the remnant of SN\,1987A}
\author[R. Wesson et al.]
{R. Wesson$^1$,  M. J. Barlow$^2$, M. Matsuura$^2$, B. Ercolano$^{3,4}$\\
$^1$European Southern Observatory, Alonso de C\'ordova 3107, Casilla 19001, Santiago, Chile\\
$^2$Department of Physics and Astronomy, University College London, Gower Street, London WC1E 6BT, UK\\
$^3$Universit\"ats-Sternwarte M\"unchen, Ludwig-Maxmilians Universit\"at M\"unchen, Scheinerstr 1, D-81679 M\"unchen, Germany\\
$^4$Excellence Cluster Universe, Boltzmannstr 2, D-85748 Garching, Germany
}
\date{Received:}

\begin{document}
\maketitle

\begin{abstract}

The discovery with the {\it Herschel Space Observatory} of bright far infrared and submm emission from the ejecta of the core collapse supernova SN\,1987A has been interpreted as indicating the presence of some 0.4--0.7\,M$_\odot$ of dust.  We have constructed radiative transfer models of the ejecta to fit optical to far-infrared observations from the literature at epochs between 615 days and 24 years after the explosion, to determine when and where this unexpectedly large amount of dust formed.

We find that the observations by day 1153 are consistent with the presence of 3$\times$10$^{-3}$M$_\odot$ of dust.  Although this is a larger amount than has previously been considered possible at this epoch, it is still very small compared to the amount present in the remnant after 24 years, and significantly higher dust masses at the earlier epochs are firmly ruled out by the observations, indicating that the majority of the dust must have formed at very late times.  By 8515-9200 days after the explosion, 0.6--0.8\,M$_\odot$ of dust is present, and dust grains with radii greater than 2\,$\mu$m are required to obtain a fit to the observed SED.  This suggests that the dust mass increase at late times was caused by accretion onto and coagulation of the dust grains formed at earlier epochs.

These findings provide further confirmation that core collapse supernovae can create large quantities of dust, and indicate that the reason for small dust masses being estimated in many cases is that the vast majority of the dust forms long after most supernovae have been detectable at mid-infrared wavelengths.

\end{abstract}
 

\section{Introduction}

The discovery that quasars at redshifts $>$5 can already contain up to 10$^{8}$M$_{\odot}$ of dust (e.g. \citealt{2003AandA...409L..47B, 2007ApJ...662..927D}) triggered vigorous research aimed at identifying the sources of the dust.  Massive stars have long been considered the most likely source due to their rapid progress to the late stages of stellar evolution where they may become significant dust producers (\citealt{1967AnAp...30.1039C}).  In particular, dust is predicted to be formed in core collapse supernovae (CCSN), in quantities sufficient to account for the dust observed in the distant universe: a number of authors have modeled dust formation processes inside CCSN ejecta, concluding that at least 0.1~M$_{\odot}$ of dust can condense (e.g. \citealt{2001MNRAS.325..726T, 2010ApJ...713....1C, 2013ApJ...776..107S}), while other investigators (e.g. \citealt{2003MNRAS.343..427M, 2007ApJ...662..927D}) have estimated that CCSN dust yields of between 0.1 and 0.5~M$_{\odot}$ per CCSN are necessary to account for the observed quantities of dust in high-z galaxies.  However, many mid-infrared studies of warm dust formation in core collapse supernova ejecta in the local universe have found that the amount of dust forming in the first few years appeared to be much smaller than required (\citealt{2007MNRAS.375..753E, 2010ApJ...715..541A, 2011ApJ...731...47A, 2012ApJ...753..109G, 2004MNRAS.352..457P, 2008ApJ...680..568S}).

Dust formation in the ejecta of CCSNe may be detected by a) a steepening of the decline in the optical brightness, as the extinction increases due to new dust formation; b) the development of an infrared excess due to optical photons being absorbed and re-emitted at longer wavelengths; and c) a blue-shifting of emission lines from the ejecta, due to the red-shifted emission from the far side of the ejecta being absorbed by dust along the line of sight.  SN 1987A was one of the first CCSNe in which dust formation was definitively detected (\citealt{1989LNP...350..164L, 1993ApJS...88..477W, 1993MNRAS.261..535M, 1993AandA...273..451B}).  Early studies assuming a smooth dust distribution estimated the presence of about 10$^{-4}$M$_{\odot}$ of dust.  However, evidence for clumping of ejecta was discussed by \citet{1991supe.conf...82L}, and \citet{2007MNRAS.375..753E} carried out an analysis using 3D radiative transfer models of clumped ejecta, finding somewhat more dust than analyses assuming a smooth distribution.  However, the amount was still far lower than theoretical predictions and, if typical of CCSNe, would mean that they could not contribute significantly to the dust budget of distant quasars.

SN\,1987A was unexpectedly detected in Herschel Space Observatory observations of the Large Magellanic Cloud (LMC) taken in 2010 (\citealt{2011Sci...333.1258M}).  The far infrared and submm emission from the position of the supernova was consistent with continuum emission from between 0.4 and 0.7 M$_{\odot}$ of dust, vastly more than was thought to have formed within a few years of the explosion.  More recent observations with the Atacama Large Millimeter Array (ALMA) confirmed that the emission was located within the expanding supernova ejecta and not in the pre-existing circumstellar material (\citealt{2014ApJ...782L...2I}), while follow-up spectroscopy using the Fourier-Transform Spectrograph of the SPIRE instrument on board {\it Herschel} showed that the SPIRE broad-band images were dominated by continuum emission, with an upper limit of 8.4\% on the contribution of emission lines to the measured broadband fluxes (\citealt{2013ApJ...773L..34K}).  Independent evidence for dust in the ejecta at these late times is provided by the asymmetry in Balmer line profiles observed by \citet{2013ApJ...768...88F} on days 7982--7998, indicating that internal dust was preferentially absorbing optical emission from the far side of the ejecta.

The recent unambiguous detections of large amounts of dust in this and other supernova remnants (e.g. \citealt{2012ApJ...760...96G}, \citealt{2010A&A...518L.138B}) show that, in at least some cases, supernovae produce quantities of dust easily sufficient to explain the quantities of dust found in distant quasars.  However, the discovery raises the question of when and how this large amount of dust formed in SN\,1987A, and whether similar amounts of dust could have formed in the ejecta of other supernovae, despite the generally small quantities measured within $\sim$5 years of the explosions for CCSNe previously studied at mid-IR wavelengths.  Are the dust masses of 10$^{-3}$--10$^{-4}$M$_\odot$ determined at these early times unreliable for any reason, with large amounts of dust forming unseen within the first few years after the outburst?  Or does the majority of the dust form at later times?

In this study we present new radiative transfer models of the ejecta of SN\,1987A at seven epochs between 3 and 24 years after its outburst.  The self-consistent models at each epoch demonstrate the physical plausibility of the large dust mass estimated 24 years after the explosion, and place strong constraints on when this dust could have formed.


\section{Observational data}

We construct models to fit the observations at seven epochs.  We use 0.3--30\,$\mu$m spectrophotometry from \citet{1993ApJS...88..477W}, obtained on days 615 and 775 after the explosion, and we supplement these data with far infrared observations from \citet{1989BAAS...21.1215H}, who observed the supernova at 100\,$\mu$m on day 626, and at both 50\,$\mu$m and 100\,$\mu$m on day 793.  We dereddened the \citet{1993ApJS...88..477W} data in the same way as described in \citet{2007MNRAS.375..753E}.

In addition to these two epochs which were modelled by \citet{2007MNRAS.375..753E}, we also created models for five subsequent epochs, 1153, 1300, 1650, 8515 and 9200 days after the explosion.  Models for the ejecta 1153 days after the explosion were constructed to match the data described in \citet{1992ApJ...389L..21D}, which consists of Kuiper Airborne Observatory spectrophotometry from 16--30\,$\mu$m, combined with photometry in filters from U to Q$_0$ taken at the Cerro Tololo Interamerican Observatory (U to I; \citet{1990AJ.....99..650S}) and the European Southern Observatory (J to Q$_0$; \citet{1993AandA...273..451B}).  We measured the fluxes plotted in Figure~1 of \citet{1992ApJ...389L..21D}, excluding the points longward of 24.5\,$\mu$m which are affected by noise and line emission, to derive a flux of 0.53$\pm$0.10\,Jy at a mean wavelength of 21.2$\pm$3.7\,$\mu$m.  Finally at this epoch we construct our model to respect an upper limit at 50\,$\mu$m of 1.65\,Jy at days 1179-83 (Harvey P., priv. comm.)

SN\,1987A was detected at a wavelength of 1.3mm with the Swedish-ESO Submillimetre Telescope (SEST) on days 1285 and 1645 (\citealt{1992AandA...255L...5B}); we supplemented these data with J-Q infrared photometry from \citet{1993AandA...273..451B} at day 1317, and radio observations at 1.4, 2.4, 4.8, and 8.6 GHz on days 1270-1306 and 1637-1662 from \citet{2010ApJ...710.1515Z}.  At these radio wavelengths the emission from the remnant is dominated by synchrotron radiation.  The synchrotron flux is observed to increase with time, while dust emission fades as the remnant expands and cools, so that synchrotron radiation may contribute significantly to longer wavelength fluxes at late times.  We therefore used the observed radio fluxes to extrapolate a power law spectrum which we then added to the model predictions for dust emission at epochs after day 1150, to obtain a total predicted flux.

Finally, we construct models to match the observations presented in \citet{2011Sci...333.1258M}, which were obtained with the Herschel Space Observatory on 30 April and 5 August 2010, days 8467 and 8564 after the explosion was observed; and newly obtained {\it Herschel} photometry from days 9090 and 9122 after the explosion (Matsuura et al., submitted) together with ALMA 440 and 870\,$\mu$m photometry obtained between days 9273 and 9387 (\citealt{2014ApJ...782L...2I}).

The observational data are summarised in Table~\ref{obs-summary}.
\begin{table*}
\begin{tabular}{llll}
\hline
Epoch &  Wavelengths &  Ref.  &  Notes \\
(days) & ($\mu$m) & & \\
\hline
615, 632, 638 & 0.3-30 & \citet{1993ApJS...88..477W}, \citet{1989Natur.340..697M} & {\em KAO} spectrophotometry \\
623-631 & 100 & \citet{1989BAAS...21.1215H} & $<$0.6 Jy \\
775 &  0.3-30 & \citet{1993ApJS...88..477W} & {\em KAO} spectrophotometry \\
791-795 & 50 & \citet{1989BAAS...21.1215H} & 2.5 $\pm$ 0.85 Jy \\
    & 100 & \citet{1989BAAS...21.1215H} & $<$1.35 Jy \\
1110 & J-Q$_0$ & \citet{1993AandA...273..451B} & ESO \\
1144 & U-I & \citet{1990AJ.....99..650S} & ESO/CTIO \\
1153 & 16-29 & \citet{1992ApJ...389L..21D} & 0.53$\pm$0.10 Jy ; {\em KAO} spectrophotometry \\
1179-1183 & 50 & P. Harvey (priv. comm.) & $<$1.65 Jy; {\em KAO} \\
1270-1306 & 1.4--4.8GHz & \citet{2010ApJ...710.1515Z} & ATCA \\
1283-1290 & 1300 & \citet{1992AandA...255L...5B} & 7.59 $\pm$ 2.45 mJy \\
1296 & V-R & \citet{1991PASP..103..958W} & CTIO \\
1317 & J-L & \citet{1993AandA...273..451B} & ESO \\
     & 10.4 (N) & \citet{1993AandA...273..451B} & $<$0.062 Jy \\
     & 18.5 (Q$_0$) & \citet{1993AandA...273..451B} & $<$0.82 Jy \\
1637-1662 & 1.4--8.6GHz & \citet{2010ApJ...710.1515Z} & ATCA \\
1644-1646 & 1300 & \citet{1992AandA...255L...5B} & 18.8 $\pm$ 4.0 mJy \\
8014 & 1.4--4.8GHz & \citet{2010ApJ...710.1515Z} & ATCA; flux densities at subsequent epochs obtained by extrapolation. \\
8467-8564 & 100--500 & \citet{2011Sci...333.1258M} & {\em Herschel} photometry \\
9090-9122 & 70--500 & Matsuura et al. (submitted) & {\em Herschel} photometry \\
9273-9387 & 450, 850 & \citet{2014ApJ...782L...2I} & ALMA band 7 \& 9 photometry \\
\hline
\end{tabular}
\caption{Summary of observation data}
\label{obs-summary}
\end{table*}

\section{Modelling}

We analyse the formation of dust in the ejecta from SN\,1987A by constructing self consistent three dimensional models, illuminated by physically plausible heating sources, using the radiative transfer code {\sc mocassin} (\citealt{2003MNRAS.340.1136E}, \citealt{2005MNRAS.362.1038E}, \citealt{2008ApJS..175..534E}).  We model the ejecta as being highly clumped, with clump sizes of between 1-3\% of the ejecta radius.  We adopt this geometry based on the expectation that Rayleigh-Taylor instabilities with characteristic size of this order will rapidly develop at the onset of the explosion (\citealt{1989ApJ...341L..63A}).  The hydrodynamical instabilities may be driven by the passage of the shockwave through density jumps in the first few thousand seconds after the explosion, or after a few days by the bursting of bubbles of gas heated by the decay of $^{56}$Ni (\citealt{1994ApJ...425..264B}).  Observational evidence for clumping in the ejecta of SN\,1987A has been discussed by \citet{1991supe.conf...82L} and \citet{1990AandA...236L..17B}.

\citet{2007MNRAS.375..753E} developed two varieties of clumpy models, the first (referred to as {\it clumpy I}) representing the case in which the dust and heating source are present in both the clumps and the interclump medium, and the second ({\it clumpy II}) in which dust is located only in the clumps, and the heating source is located only in the interclump medium.  In this work we consider only the clumpy II scenario; the dust mass that can be contained in optically thick clumps is maximised in this scenario, and thus we can place an upper limit on the amount of dust formed at each epoch prior to the Herschel observations, and thus determine the minimum amount of additional dust which must have formed during the 20 years between our third and fourth epochs.  In our models, clumps are distributed stochastically with a number density proportional to r$^{-2}$, and the clumps themselves have a $r^{-2}$ density profile, modelled using a 5x5x5 subgrid inside the mother grid cell in which the clump is located.  The number of clumps in the model is specified via a volume filling factor, $f$, which represents the volume of the expanding shell occupied by the clumps.  We investigated volume filling factors between $f$=0.025 and 0.2.  We also considered clump number densities proportional to r$^0$ and r$^{-4}$ where r is the radius of the ejecta; the predicted optical and infrared peaks of the SED vary by $\pm\sim$20\%, with a steeper density profile resulting in a higher optical flux and lower infrared flux.  The predicted SED was found to be insensitive to the density distribution in the clumps themselves.

We investigated various mixtures of amorphous carbon and silicate dust, using optical constants from \citet{1984ApJ...285...89D} for silicates and \citet{1988ioch.rept...22H} for amorphous carbon.  For all our models we adopted a distance to the LMC of 49.97 kiloparsecs (\citealt{2013Natur.495...76P}).  A list of all the model parameters is given in Table~\ref{modelparameters}

\begin{table*}
\begin{tabular}{ll}
\hline
Model parameter & Parameter space \\
\hline
Inner radius                       & 5$\times$10$^{15}$cm (day 615) -- 7.5$\times$10$^{16}$cm (day 9200) \\
Outer radius/inner radius          & 5 \\
Clump radius                       & R$_{\rm out}$/30, R$_{\rm out}$/45, R$_{\rm out}$/60 \\
Clump number density               & n$_{\rm clump}$(r)\,$\propto$\,r$^{-0,2,4}$ \\
Clump density profile              & n$_{\rm dust}$\,$\propto$\,r$^{-0,2,4}$ \\
Clump volume filling factor        & 0.025--0.2 \\
Luminosity of heating source       & 200--5.7$\times$10$^5$\,L$_\odot$ \\
Temperature of heating source      & 3000--7000\,K \\
Spectrum of heating source         & Blackbody \\
Dust mass                          & 10$^{-4}$--2.0\,M$_\odot$ \\
~~~~Fraction in interclump medium  & 0\% \\
~~~~Fraction in clumps             & 100\% \\
Dust species                       & Amorphous carbon, astronomical silicates, iron \\
Distribution of grain radii        & n(a)$\propto$(a-x)$^{-3.5}$, with (x+0.005)$<$a$<$(x+0.25) $\mu$m; x=0,1,2,3,4,5 $\mu$m \\
\hline
\end{tabular}
\caption{Summary of all model input parameters and the ranges of values investigated.  Not all possible combinations of parameters were investigated, the range depending on the epoch being modelled.}
\label{modelparameters}
\end{table*}

We began by developing viable models for day 615.  We then created models for subsequent epochs by evolving the day 615 models assuming that all parts of the ejecta expanded with constant velocity between epochs.  Thus, the inner edge of the ejecta is expanding at 941\,kms$^{-1}$.  We then varied the luminosity and dust parameters to achieve a fit to the observations of the subsequent epoch.  The primary dust variable is the mass; if varying the dust mass could not give a good fit, we also varied the dust composition and the grain size distribution.

The dust heating source at the first three epochs is the radioactive decay of $^{56}$Co, with contributions from $^{44}$Ti and other radionuclides after day $\sim$900.  The luminosity used in the models for these epochs was taken from the bolometric light curve of \citet{1991AJ....102.1118S} which covers days 254--1352.

By day 8515, the decay of $^{44}$Ti is the dominant heating source, with possible additional contributions from X-rays produced in shock interactions between the ejecta and pre-existing circumstellar material, and the diffuse interstellar field.  We discuss the heating luminosity at this epoch in Section~\ref{sectiond8515}.

\subsection{Reproducing previous results}

Using mocassin v2.02.70, we first attempted to reproduce the results of \citet{2007MNRAS.375..753E}.  In an erratum, \citet{2007MNRAS.379.1248E} noted that mocassin version 2.02.21, which they had used to calculate all their models, was afflicted by a bug which affected the calculation of the IR SED.  Probability density functions used for calculating dust emission were incorrect, while those used for absorption calculations were unaffected.  This meant that while the shape of the calculated IR excess was incorrect, the optical SED was unaffected.

\begin{figure*}
\includegraphics[width=0.47\textwidth]{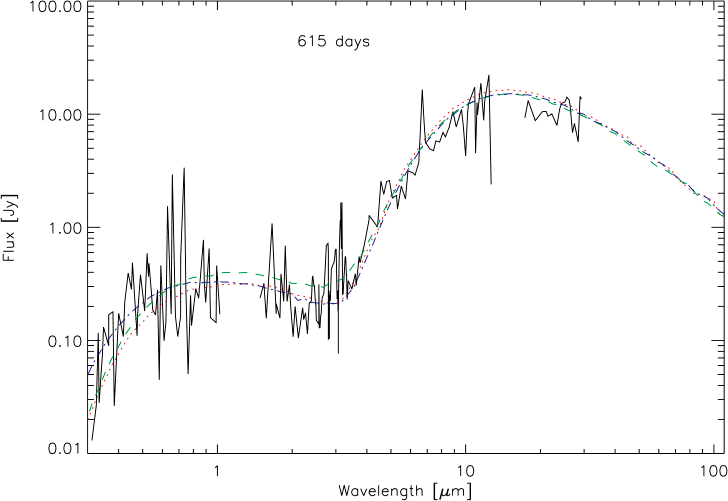}
\includegraphics[width=0.47\textwidth]{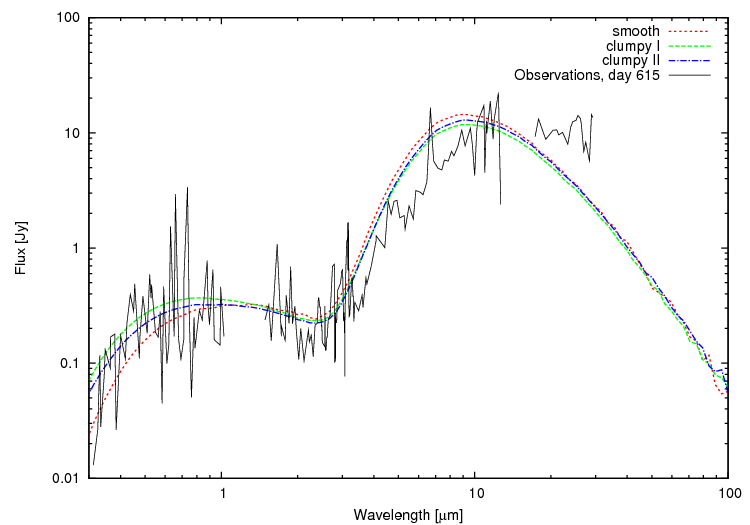}
\caption{(l) Figure 8 from \citet{2007MNRAS.375..753E}, showing model SEDs created using mocassin v2.02.21;
(r) SEDs created using parameters from Table 1 of \citet{2007MNRAS.375..753E} and mocassin v2.02.70}
\label{reproductions}
\end{figure*}

The bug was fixed with the release of v2.02.41 of mocassin; we recalculated the models presented in \citet{2007MNRAS.375..753E} using mocassin v2.02.70.  Figure~\ref{reproductions} shows the results of model calculations for amorphous carbon, created using the input parameters specified in Table 1 of \citet{2007MNRAS.375..753E}, compared to the output obtained with mocassin v2.02.21.  The optical portion of the reproduced SED is very similar to the original, while the shape of the IR SED differs, as expected from the nature of the bug.  Figure~\ref{reproductions} shows that the reproduced models give higher mid-infrared fluxes than the original models, and the shape of the SED significantly diverges from the original at wavelengths longer than about 10\,$\mu$m.  The lower fluxes at longer wavelengths allow the possibility that a higher dust mass than previously reported could be necessary to fit the data.  We therefore began by searching for ejecta parameters which better fit the data at day 615.

\subsection{Day 615}
\label{day615section}

Using the models originally presented in \citet{2007MNRAS.375..753E} as a basis, we varied the model parameters to regain a good fit to the IR SED.  We used the same R$_{in}$ and Y(=R$_{out}$/R$_{in}$) as adopted by \citet{2007MNRAS.375..753E} (5.0e15\,cm and 5 respectively), as well as the same grain size distribution, with the number density of dust particles with a radius $a$ being proportional to $a^{-3.5}$ between limits of 0.005 and 0.25\,$\mu$m.  We then ran models for volume filling factors, $f$, of 0.025, 0.05, 0.1 and 0.2, and for dust masses between 10$^{-4}$ and 1.0\,M$_\odot$.  At this epoch, we investigated different clumping scales, creating models with clump radii of R/30, R/45 and R/60.  Varying the clump size did not have a major effect on the SED.  The number of grid cells required for a clumping scale of R/x is proportional to x$^{3}$ and so the computing time required becomes prohibitively expensive as the clump size is decreased; at subsequent epochs we therefore modelled ejecta only with a clump size of R/30.  This size is in agreement with theoretical predictions of the characteristic scale of the Rayleigh-Taylor instabilities predicted to form in supernova ejecta (eg \citealt{1994ApJ...425..814H}).

We found that an improved fit to the observed day 615 SED could be obtained by increasing the dust mass in the \citet{2007MNRAS.375..753E} models.  The volume filling factor {\it f} is well constrained - as described in \citet{2007MNRAS.375..753E}, for optically thick clumps in the {\it clumpy II} scenario, the emergent optical flux is determined solely by {\it f}, and as noted in \citet{2007MNRAS.379.1248E}, the optical flux in the models produced with mocassin 2.02.21 was not affected by the bug in that version.  Therefore it remains the case that f=0.1 provides the best fit to the optical SED.  Fits to the IR SED are then obtained by varying the dust mass only.

\subsubsection{The carbon and silicate fractions}

\begin{figure}
\includegraphics[width=0.47\textwidth]{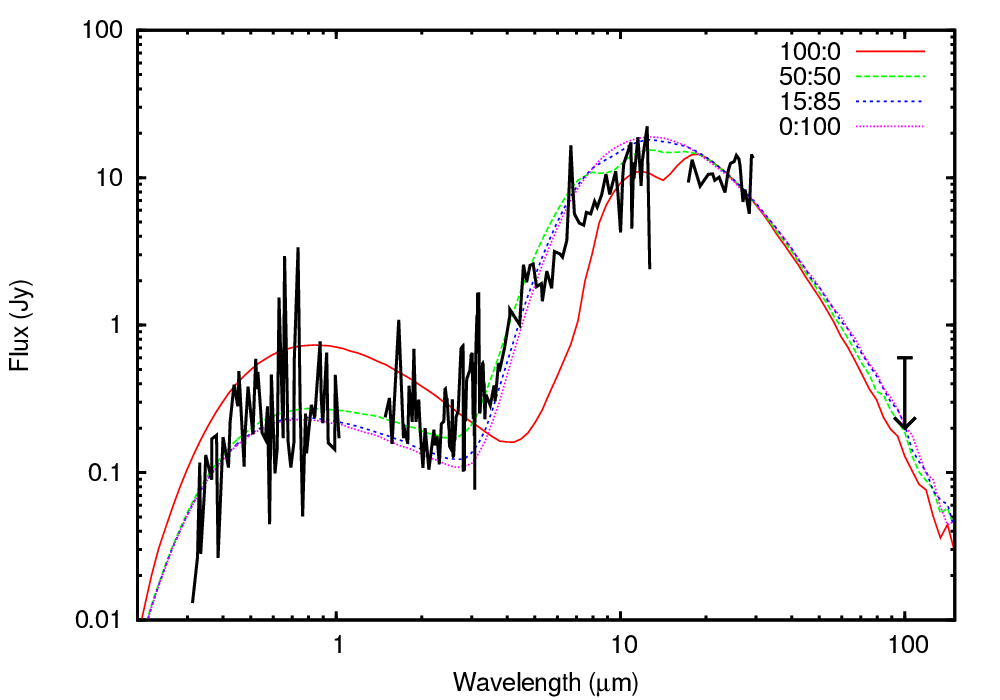}
\caption{Effect of dust composition on the predicted SED at day 615 with all other parameters including total mass held constant; the ratios are silicates:carbon by mass}
\label{day615_composition}
\end{figure}

We investigated dust compositions at this epoch of pure carbon, pure silicates, and mixtures of silicates:carbon of 15:85 and 50:50.  Plots of the predicted SED at day 615 for a dust mass of 0.001\,M$_\odot$ with each composition are shown in Figure~\ref{day615_composition}.  As can be seen in the figure, pure silicate models are ruled out by the data; at dust masses of $\sim$0.001\,M$_\odot$, a reasonable fit to the 10--100\,$\mu$m region of the SED is obtainable, but the emission between 3--10\,$\mu$m is strongly underpredicted.  Increasing the dust mass to 0.01\,M$_\odot$ results in an improved fit to the shorter wavelength region, but the 10-100\,$\mu$m emission is then strongly overpredicted.

Models with a 50:50 mixture of silicates and amorphous carbon provide a somewhat better fit to the data, but cannot fit all parts of the IR excess; lower dust masses result in an underpredicted 3--10\,$\mu$m region while higher dust masses overpredict the 10--100\,$\mu$m region.  Any dust mass higher than 0.01\,M$_\odot$ overpredicts the 100\,$\mu$m observation.

\begin{figure}
\includegraphics[width=0.47\textwidth]{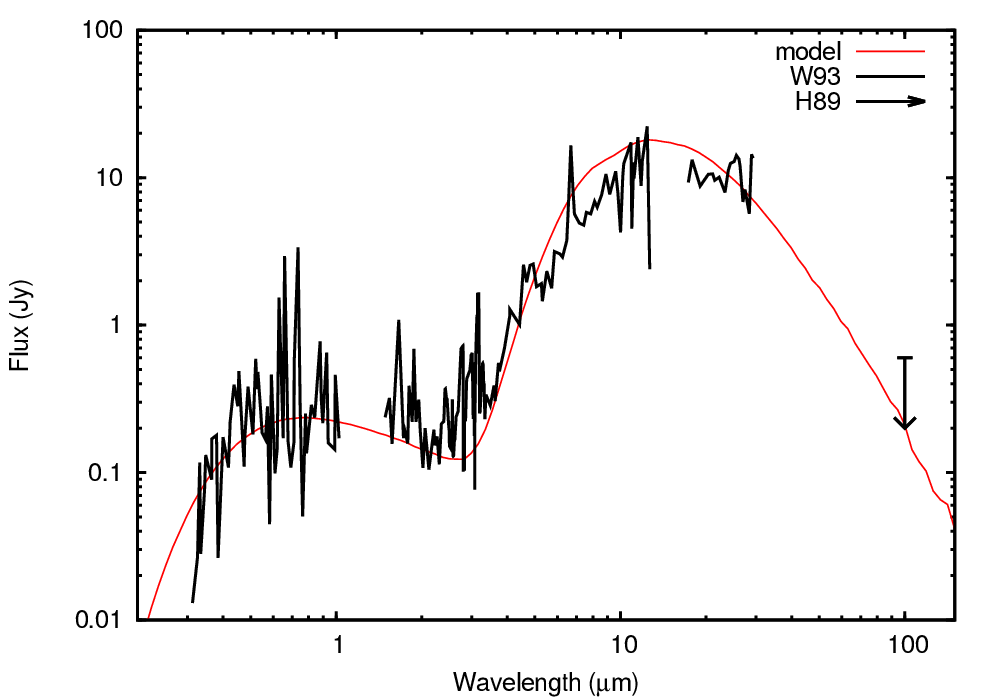}
\caption{Best fitting model for day 615: M=0.001\,M$_\odot$, f=0.1, 85:15 carbon:silicate ratio, 0.005\,$\mu$m$<$a$<$0.25\,$\mu$m.  Optical-IR spectrophotometry is from Wooden et al. (1993) and the 100\,$\mu$m photometry is from Harvey et al. (1989).}
\label{day615figure}
\end{figure}

Pure carbon models can reproduce all parts of the SED well.  Models with a 15:85 ratio of silicates to carbon provide a slightly improved fit to the observations at 2-4\,$\mu$m.   Figure~\ref{day615figure} shows our best fitting model at day 615.  The model has a volume filling factor of 0.1 and a clump size of R/30.  The dust mass is 1.0$\times$10$^{-3}$M$_{\odot}$, approximately a factor of 5 higher than that found by \citet{2007MNRAS.375..753E}; the best fitting amorphous carbon model at day 615 in \citet{2007MNRAS.375..753E} had a dust mass of 2.2$\times$10$^{-4}$M$_{\odot}$.  From our models with slightly higher and lower dust masses compared to the best fitting model, we estimate an uncertainty on our derived dust mass of $\pm$2.0$\times$10$^{-4}$M$_\odot$.

\begin{figure}
\includegraphics[width=0.47\textwidth]{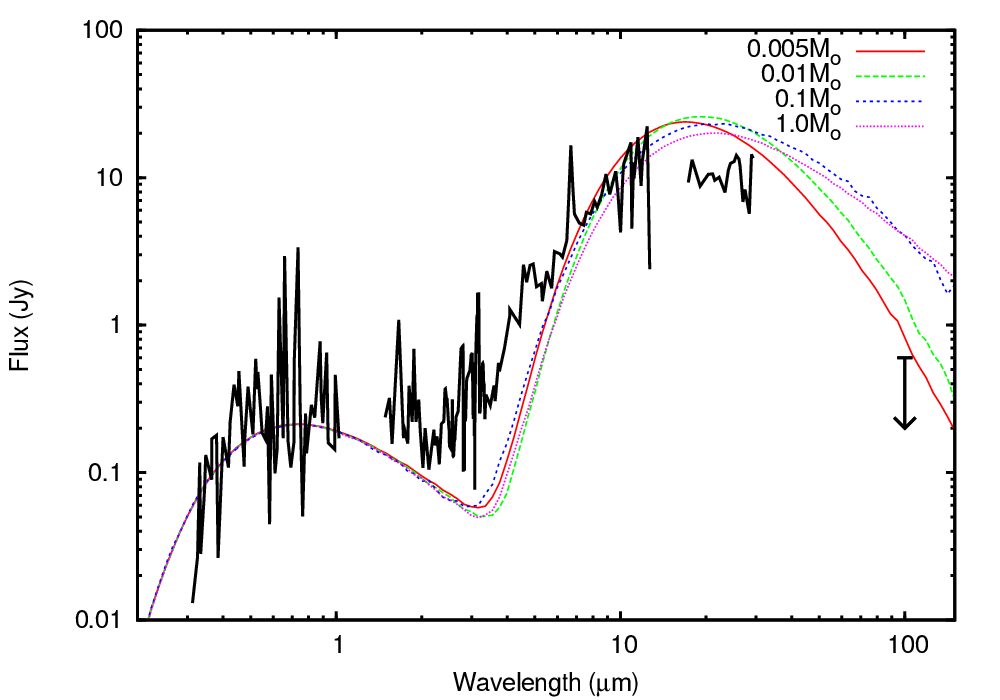}
\caption{Predicted SEDs at day 615 for dust masses higher than 0.005\,M$_\odot$, f=0.1, 85:15 carbon:silicate ratio, 0.005\,$\mu$m$<$a$<$0.25\,$\mu$m}
\label{day615_highmassruledout}
\end{figure}

Regardless of dust composition, the day 626 upper limit at 100\,$\mu$m provides a strong constraint on the dust mass, requiring that it be lower than 0.01\,M$_\odot$ in all cases.  For the 15:85 composition which provides the best overall fit to the SED, the 100\,$\mu$m flux upper limit of 0.6~Jy (\citealt{1989BAAS...21.1215H}) is incompatible with any dust mass higher than 0.005\,M$_\odot$ (Figure~\ref{day615_highmassruledout}).

\subsection{Day 775}

To fit the observations at day 775, we radially expanded the grid for the best fitting day 615 model by a factor of (775/615) and then varied the dust mass to achieve a fit.  We find that this alone enabled a reasonable fit to the day 775 data, with no need to vary the volume filling factor.  The dust mass required to match the data is 2.0$\times$10$^{-3}$M$_{\odot}$, consistent with further dust formation having occurred between days 615 and day 775, in the same region in which it had formed by day 615.  As with our estimated day 615 dust mass, this mass is approximately a factor of 5 higher than found by \citet{2007MNRAS.375..753E} at this epoch with the same dust composition.

\begin{figure}
\includegraphics[width=0.47\textwidth]{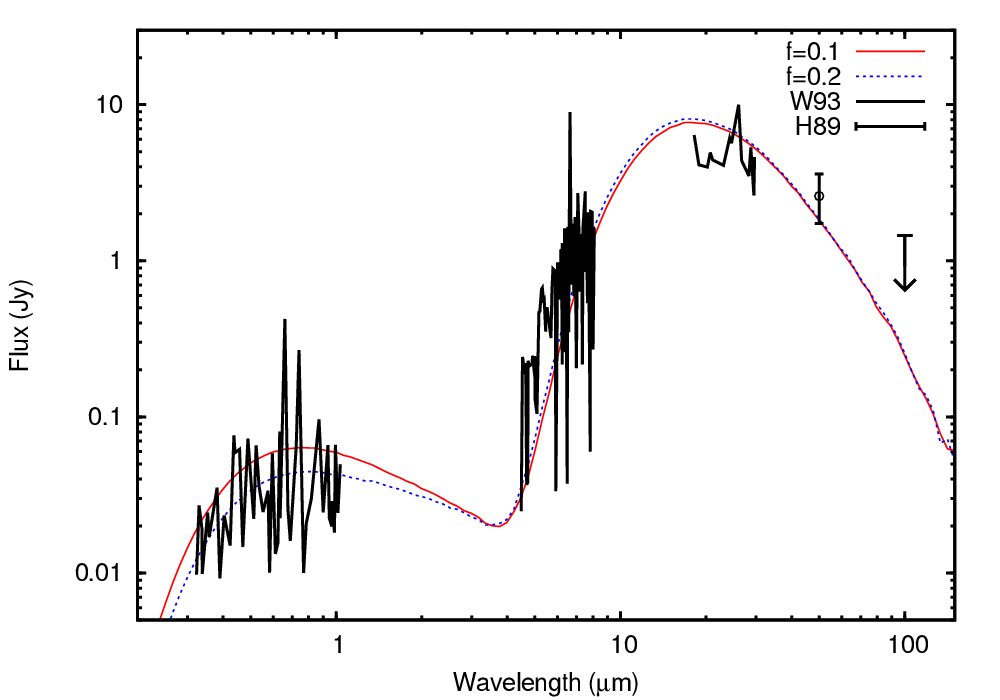}
\caption{Best fitting models for day 775: M=0.002\,M$_\odot$, 15:85 silicates:carbon, 0.005\,$\mu$m$<$a$<$0.25\,$\mu$m.  Models with volume filling factors of f=0.1 and f=0.2 are shown.}
\label{day775figure}
\end{figure}

A slightly better fit to the optical SED is achieved with a volume filling factor of f=0.2 (see Figure~\ref{day775figure}).  This could suggest that between days 615 and 775, dust formed in clumps which had previously been dust free.  However, models with the same value of f as in day 615 are not ruled out and thus it is also possible that the extra dust mass came from further dust forming in the same clumps in which it had previously been present.

\subsection{Dust mass limits at early epochs}

With the discovery of the presence of large amounts of dust in the remnant of SN\,1987A 23 years after the explosion, it is of great importance to determine when this large amount of dust formed.  Studies which have quantified the amount of dust present in young ($<$3 years old) supernova remnants including SN\,1987A have typically found dust masses of 0.001\,M$_\odot$ or less.  This raises the question of whether those low dust masses might have been underestimates.

In this study we find that in our best fitting model of the ejecta at day 615, the infared emission at 10\,$\mu$m is optically thick and insensitive to an increase in the total mass of dust for dust masses greater than about 0.001\,M$_\odot$.  However, the far infrared data points at days 615 and 775 provide strong upper limits on the dust mass that could be hidden inside dense clumps at these epochs, as shown in Figure~\ref{day615_highmassruledout}.  To obtain a flux below the 3$\sigma$ upper limit of 0.6\,Jy at 100\,$\mu$m on day 615 with a dust mass as high as 1$M_\odot$ would require that the volume occupied by the clumps be very small; the emission at 100\,$\mu$m is still optically thin with f=0.025 and M=0.01\,M$_\odot$, and already exceeds the observed flux by a factor of 3.5.  But models with volume filling factors below 0.025 significantly overpredict the optical fluxes.  Models in which the volume filling factor gives optical fluxes consistent with the observations place a strong upper limit of $\sim$0.002\,M$_\odot$ on the dust mass at day 615.  Figure~\ref{day615_highmassruledout} shows the model outputs for a volume filling factor of 0.1 and dust masses of 0.005, 0.01, 0.1 and 1.0\,M$_\odot$ at day 615, showing that the emission at 100\,$\mu$m becomes optically thick only for dust masses greater than 0.01\,M$_\odot$, where the emission at 100\,$\mu$m already significantly exceeds the observed value.

\subsection{Day 1153}

To model the day 1153 observations we proceeded as before, expanding the day 615 grid by a factor of (1153/615) and varying the dust mass.  The uncertainties on the L and M band fluxes reported by \citet{1993AandA...273..451B} imply that the detections were significant only at the level of 2.5 and 1.6$\sigma$ respectively, and we therefore use 3$\sigma$ upper limits as constraints on our models.  In their analysis of the day 1153 observations of SN\,1987A, \citet{1992ApJ...389L..21D} took N and Q band fluxes from \citet{1993AandA...273..451B}, obtained at ESO, and rescaled them to match photometry obtained at the CTIO.  Without this rescaling, we find that the measured fluxes cannot be matched by any model, with no combination of dust mass and luminosity predicting a flux as high as those observed.  We therefore use the rescaled fluxes.  The systematic uncertainties are thus likely to be considerably larger than the uncertainties reported by \citet{1993AandA...273..451B}.

\begin{figure}
\includegraphics[width=0.47\textwidth]{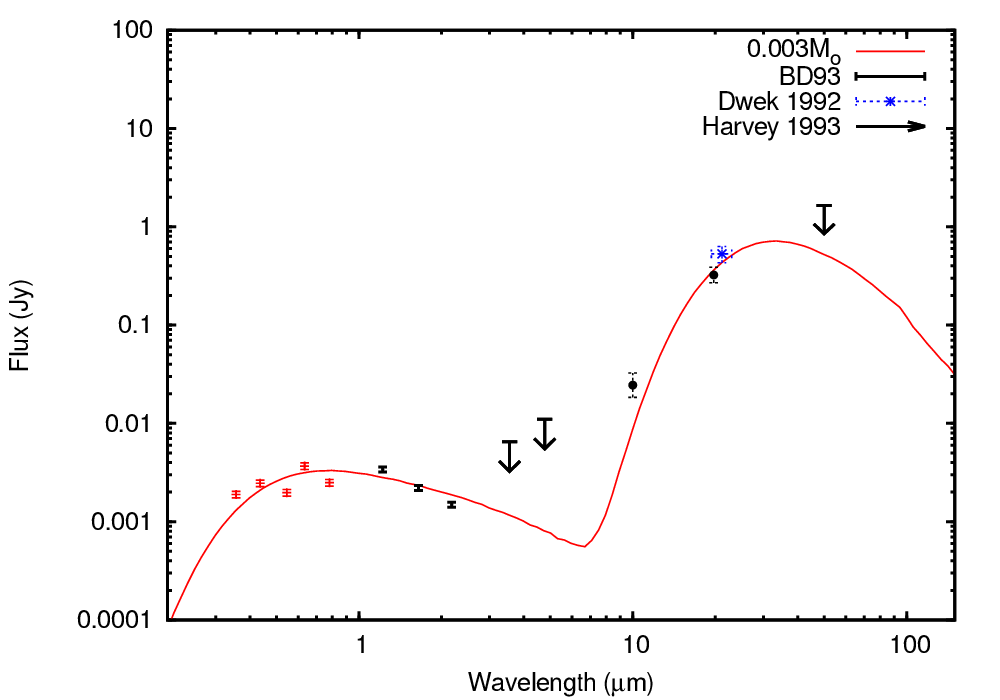}
\caption{Best fitting model for day 1153: M=0.003\,M$_\odot$, f=0.1, pure carbon, 0.005\,$\mu$m$<$a$<$0.25\,$\mu$m.  Data labelled ``BD93'' are from \citet{1993AandA...273..451B}}
\label{day1153figure}
\end{figure}

We find that a uniformly expanded grid with a dust mass of 0.003\,M$_\odot$ gives the best overall match to the SED (Figure~\ref{day1153figure}), although it underpredicts the N-band flux by approximately a factor of three.  A lower dust mass results in a better fit to this point, with the predicted SED for M$_d$=0.001\,M$_\odot$ matching the observation well; however, the optical fluxes are then overpredicted significantly, and the dust mass is lower than found at the previous epoch.  Increasing the dust mass results in cooler dust, and the N-band flux is further underpredicted.

We investigated whether varying the dust composition could improve the overall fit to the SED.  While at the two previous epochs, the shape of the SED requires that the dust be predominantly carbon, the decrease of silicon fine structure emission line strengths as well as the disappearance of iron fine structure emission have been attributed to the formation of silicate and iron dust at later epochs (\citealt{1992ApJ...389L..21D}).  The decrease in the fluxes of forbidden lines may also be due simply to the declining temperature of the gas in the expanding remnant; to investigate whether the observed SEDs could be consistent with dust of different species forming, we created models in which part of the dust was in the form of iron or silicates.  A moderate silicate fraction is possible at this epoch, but the lower opacity of silicate dust in the optical compared to carbon dust means that when the silicate fraction is higher than 25\%, the dust mass required to match the infrared SED results in the optical SED being overpredicted.  Increasing the dust mass to provide sufficient opacity to match the optical SED results in an overprediction of the infrared fluxes.  The silicate fraction at day 1153 must therefore still be much lower than the carbon fraction.

\begin{figure}
\includegraphics[width=0.47\textwidth]{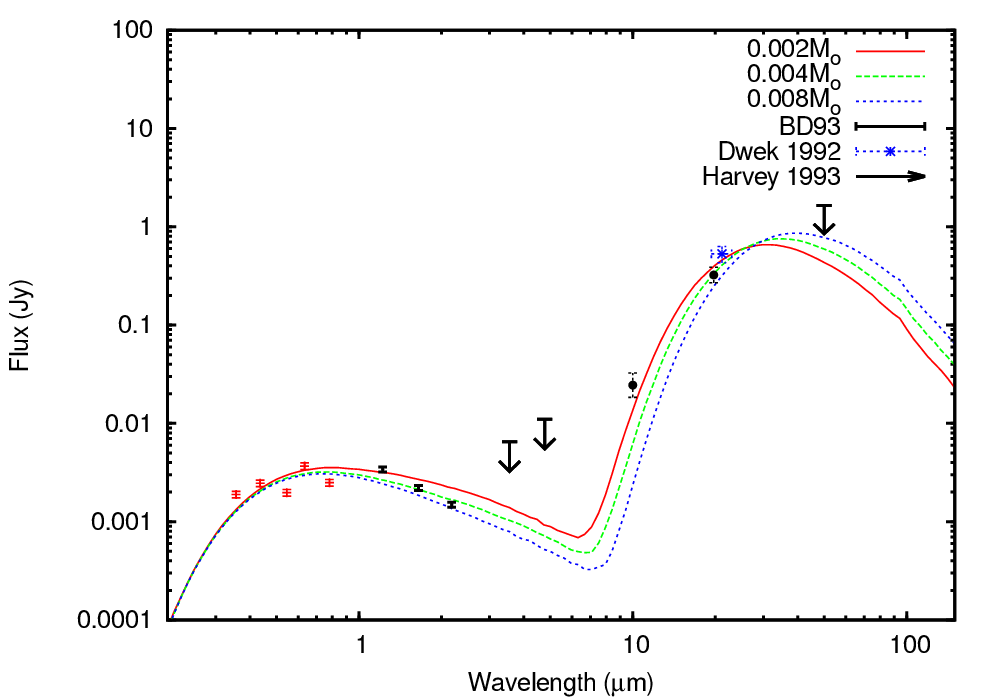}
\caption{Day 1153 models with 0.002\,M$_\odot < M_d < 0.008M_\odot$ of amorphous carbon dust}
\label{day1153_highmassruledout}
\end{figure}

The mid-infrared fluxes at this epoch place strong constraints on the amount of dust that could be present.  Figure \ref{day1153_highmassruledout} shows the predicted SED for dust masses between 0.002 and 0.008\,M$_\odot$, in the clumpy shell consistent with expansion from previous epochs.  As with the upper limit to the dust mass at earlier epochs, the dependence of the predicted optical flux on the covering factor of the dusty clumps constrains that parameter to a value inconsistent with optically thick clumps at long wavelengths.  The dust mass is thus strongly limited to less than 0.006\,M$_\odot$ at day 1153, and we consider the most likely value to be 0.003$\pm$0.001\,M$_\odot$.


\subsection{Day 1300}
\label{d1300}

The remnant of SN\,1987A was detected at a wavelength of 1.3mm, 1300 days after the explosion (\citealt{1992AandA...255L...5B}, with a measured flux was 7.59$\pm$2.45\,mJy.  Models constructed by expanding the previous grids at constant velocity and increasing the dust mass cannot satisfactorily fit this point, if it is assumed that all of the 1.3mm flux arises from dust continuum emission; even for a dust mass of 0.5\,M$_\odot$, the flux at 1.3mm is still underpredicted by a factor of $\sim$7.
Besides dust continuum emission, synchrotron radiation and free-free emission may also contribute to the flux at 1.3mm.  Using radio observations of the remnant from \citet{2010ApJ...710.1515Z} at 1.4--5.6GHz at which frequencies the emission is dominated by synchrotron emission, we estimate a synchrotron flux in Janskys at a wavelength $\lambda$ in microns given by F$_\nu$($\lambda$)=0.00138$\times (\frac{\lambda}{10^5})^{0.900}$\,Jy.  From this we find that the predicted synchrotron flux at 1.3mm is 0.03\,mJy.

\begin{figure}
\includegraphics[width=0.47\textwidth]{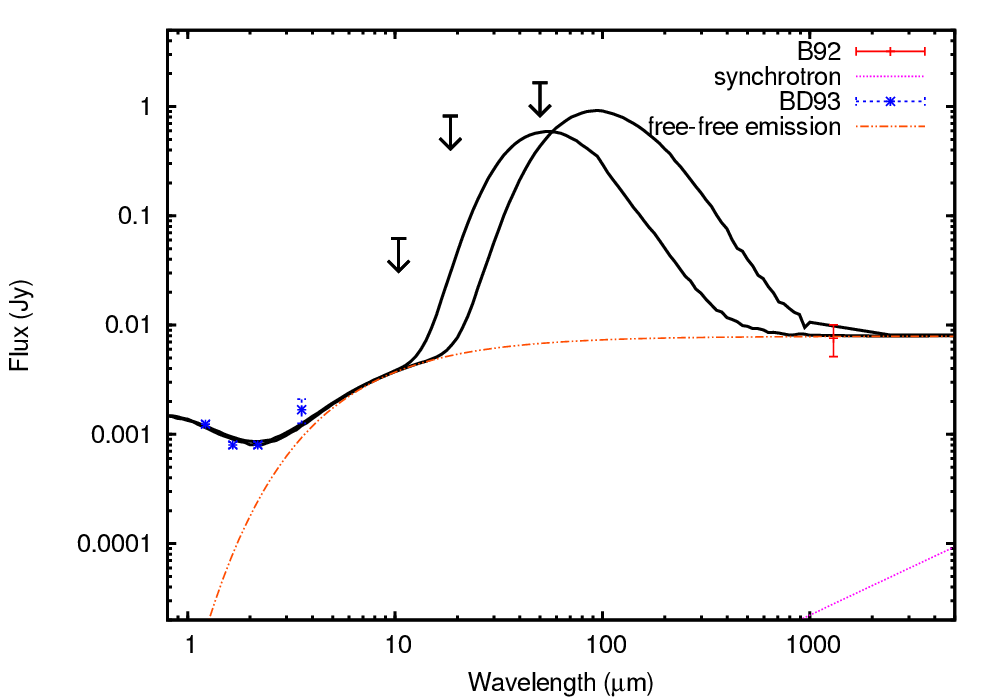}
\caption{Models at day 1300: the SED is the sum of free-free, synchrotron and dust emission.  SEDs due to 0.02 and 0.5 M$_\odot$ of amorphous carbon dust are plotted; these are the lowest and highest masses that can fulfil the observational constraints.  f=0.1 and 0.005\,$\mu$m$<$a$<$0.25\,$\mu$m for these models.  ``B92'' indicates data from \citet{1992AandA...255L...5B}.}

\label{day1300figure}
\end{figure}

To estimate the free-free emission at day 1300, we use the following expression adapted from \citet{1989PhDT........13W}:

\begin{equation}
F_{\nu,ff}(\lambda) = 5.733\times10^{11} (1-\xi)^2 {\rm exp}(\frac{-10^6hc}{\lambda kT}) T^{-0.5}t^{-3} Jy
\end{equation}

where 1-$\xi$ is the ionised fraction of the gas, T is the gas temperatured, $\lambda$ is the wavelength in microns and t is the time in days since the supernova explosion.  \citet{1989PhDT........13W} estimated the free-free flux as coming from a partially ionised 10\,M$_\odot$ hydrogen envelope.  The 3.6--1300\,$\mu$m SED can be well fitted by assuming a gas temperature of 1800\,K and an ionised fraction of (1-$\xi$)=0.037; in particular this reproduces the observation that the L band flux is considerably higher than the JHK fluxes.  The 1.3mm flux is then dominated by free-free emission, and the amount of dust is poorly constrained.  To fit the near-infrared observations, a dust mass of at least 0.02\,M$_\odot$ is required; lower dust masses do not give enough extinction and the JHK fluxes are overpredicted.  The 1.3mm flux constrains the dust mass to less than 0.5\,M$_\odot$, for the case of amorphous carbon.  We therefore find that at day 1300, 0.02$<$M$_d <$0.5\,M$_\odot$ (Figure~\ref{day1300figure}).  Our estimated lower limit is already a factor of five higher than the 0.004\,M$_\odot$ estimated on day 1153.

\subsection{Day 1650}

\begin{figure}
\includegraphics[width=0.47\textwidth]{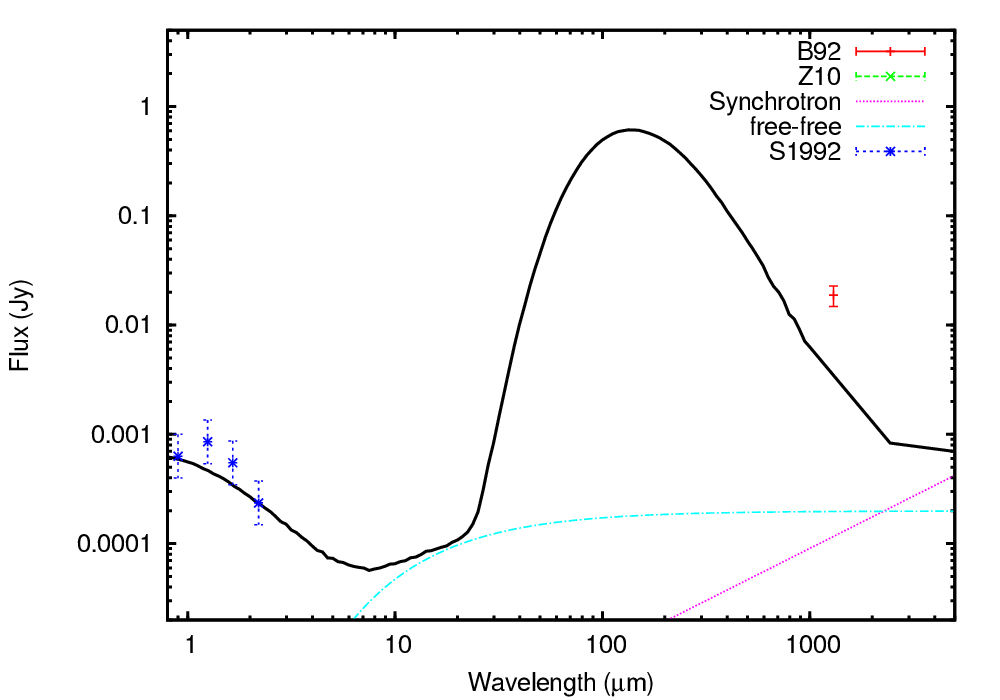}
\caption{Predicted SED for a model with 1.0\,M$_\odot$ of amorphous carbon dust on day 1650 showing that the 1.3mm flux is still underpredicted even for implausibly high dust masses.  Fluxes marked ``S1997'' are from \citet{1997astro.ph..7324S}.}
\label{day1650figure}
\end{figure}

SN\,1987A was detected again at 1.3mm in observations obtained on days 1644-1646 by \citet{1992AandA...255L...5B}, with a measured flux of 18.8$\pm$4\,mJy.  As at day 1300, we used radio data from \citet{2010ApJ...710.1515Z} to estimate the synchrotron emission at shorter wavelengths; we find that F$_\nu$($\lambda$)=0.00741$\times (\frac{\lambda}{10^5})^{0.957}$Jy, giving a predicted synchrotron flux at 1.3mm of 0.11\,mJy.  We also consider the likely free-free flux at this epoch: the free-free spectrum is flat between 1.3mm and the lower frequencies observed by \citet{2010ApJ...710.1515Z}, and so the free-free flux is limited to less than 2.5\,mJy, the flux reported at 8.6GHz.  The flux that could be due to dust continuum emission is then 16.2$\pm$4\,mJy.  This is considerably higher than the fluxes predicted by any of our models even under extreme assumptions; for a dust mass of 1.0\,M$_\odot$ consisting of dust grains as large as 3.25\,$\mu$m in radius, the predicted dust continuum emission at 1.3mm is only 4.5\,mJy (Figure~\ref{day1650figure}).  We conclude that the day 1644..1646 1300\,$\mu$m detection reported by \citet{1992AandA...255L...5B} could not have been due to dust or free-free emission.  A similar conclusion may then apply to their 1300\,$\mu$m detection on day 1300 (Section~\ref{d1300}),

\subsection{Day 8515}

The discovery by \citet{2011Sci...333.1258M} of far-IR to submm emission at the position of SN 1987A on days 8467 and 8564 was unexpected.  The emission was interpreted as being dust continuum emission, and analytical fits to the observed SED gave an estimated dust mass of 0.4--0.7 M$_\odot$, depending on the composition of the dust.  The 440- and 870-$\mu$m ALMA observations of \citet{2014ApJ...782L...2I} between days 9273 and 9387 confirmed the presence of at least 0.2\,M$_\odot$ of dust in the inner ejecta.  Given the nucleosynthesis predictions for the estimated progenitor mass of $\sim$19\,M$_\odot$ (\citealt{1990ApJ...349..222T}, \citealt{1988ApJ...330..218W}), this implies that refractory elements in the ejecta condense into dust with an efficiency approaching 100\%.

Subsequent observations with the SPIRE spectrometer on board Herschel have confirmed that the emission is indeed continuum emission.  CO 6-5 and 7-6 line emission was detected with SPIRE, and the observations place an upper limit to the line contribution to the broadband fluxes of 8.4\%.  At the same time, line and continuum observations using ALMA at high spatial resolution confirm that this emission arises from within the ejecta, definitively ruling out any thermal echos from pre-existing structures (\citealt{2013ApJ...773L..34K, 2014ApJ...782L...2I}).  The presence within the ejecta of a large amount of dust is thus confirmed.  By considering our models at earlier epochs and expanding them appropriately, we can now determine whether the high dust mass estimated by \citet{2011Sci...333.1258M} is consistent with our own modelling of the evolution of the shell, and furthermore place constraints on when and where this dust formed.

\subsubsection{The heating source}
\label{sectiond8515}

In previous epochs, the dust heating source was assumed to be from the radioactive decay of short-lived radioisotopes including $^{56}$Co.  By 24 years after the explosion, much longer-lived isotopes, primarily $^{44}$Ti, dominate the radioactive heating, and other energy sources may be significant; additional heating is expected from the X-rays emitted by the interaction between supernova ejecta and circumstellar material, and from the ambient interstellar UV field.

The luminosity expected from the decay of $^{44}$Ti is approximately 400\,L$_{\odot}$ (\citealt{2011AandA...530A..45J}).  The X-ray luminosity from ejecta-CSM interaction is approximately 500\,L$_{\odot}$, although little of this X-ray emission is expected to go into heating the newly formed dust (\citealt{2010AandA...515A...5S}).  We constructed models for day 8515 using luminosities between 225 and 900\,L$_{\odot}$.  The infrared SED is insensitive to the assumed temperature of the heating source -- models with heating source temperatures of 7000\,K (used at previous epochs) and 3000\,K predicted different optical SEDs but very similar IR SEDs.

\subsubsection{Dust grain size distribution}

To model the emission at day 8515, we expanded the day 615 grid as before, this time by a factor of (8515/615), and then varied other parameters to achieve a fit.  \citet{2010ApJ...710.1515Z} presented radio observations up to day 8014; from the exponential fitting parameters in their Table 2, we extrapolated to day 8515, and find that the synchrotron flux at this epoch in Jy at a wavelength $\lambda$ is given by F$_\nu$($\lambda$)=0.315$\times (\frac{\lambda}{10^5})^{0.635}$\,Jy.  This predicts a synchrotron flux at 500\,$\mu$m of 11\,mJy, compared to the upper limit from the Herschel observations of 29\,mJy, and 9\,mJy at 350\,$\mu$m compared to the observed flux of 49\,mJy.  The synchrotron flux is thus small but not negligible and we subtracted it from the fluxes tabulated in \citet{2011Sci...333.1258M} before proceeding.

The volume filling factor is not uniquely constrained by the observations at this epoch alone; when the SED is fitted with a given combination of $f$ and $L$, a similar fit to the observations can be obtained by increasing the volume filling factor while reducing the luminosity or vice versa.  Given that at each of the previous epochs, the volume filling factor was constrained by the observations to be 0.1, we set it to this value at day 8515.

Nucleosynthesis models appropriate for SN\,1987A predict that if dust condensation from refractory elements were 100\% efficient, then the maximum amount of dust that could be present in the remnant would be around 0.7\,M$_\odot$ (\citealt{1990ApJ...349..222T,1988PASAu...7..355W}).  When all parameters are kept the same as at previous epochs except for the luminosity and the dust mass, we are unable to fit the data well with any physically plausible combination of the two; the predicted dust SED was always warmer than observed unless the dust mass is greater than 1.0\,M$_\odot$.  We therefore considered the possibility that the large increase in dust mass by day 8515 was due to previously existing grains growing rather than new grains condensing.

It can be shown that the incremental growth of a grain by gas accretion, from a radius a$_1$ to a larger radius a$_2$, is independent of the intial radius a$_1$, so that as accretion progresses, the grain size distribution will narrow if no fragmentation is also occurring.  We therefore considered models which assume only gas accretion, linearly increasing by the same amount the size of all grains in the distribution used at earlier epochs, as well as models in which smaller dust grains remain present, either due to the formation of new dust grains, or the fragmentation of existing large grains.

\begin{figure}
\includegraphics[width=0.47\textwidth]{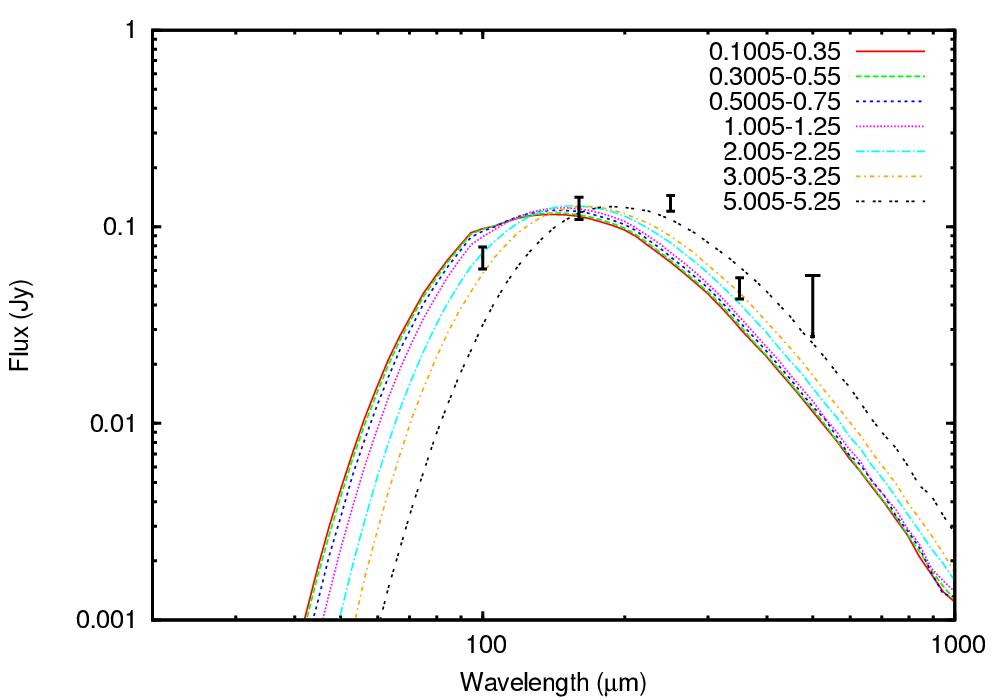}
\caption{The effect of varying the amorphous carbon dust size distribution on the fit to observed fluxes at day 8515; dust mass=0.8\,M$_\odot$, optical constants from \citet{1988ioch.rept...22H}}
\label{dustsizeeffect}
\end{figure}

\begin{figure}
\includegraphics[width=0.47\textwidth]{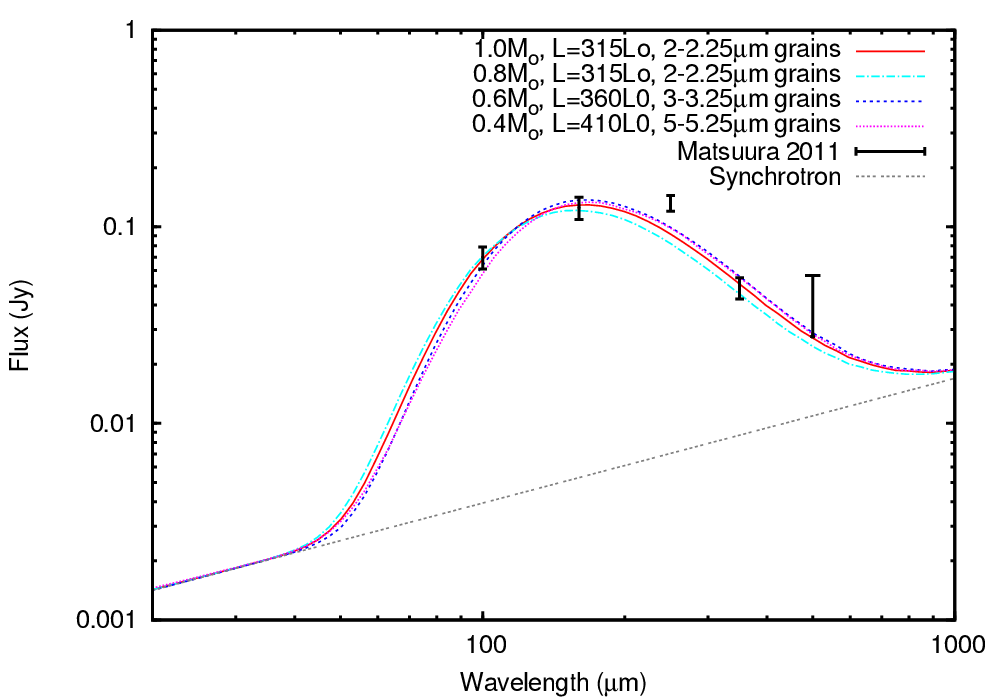}
\caption{Best fitting models for day 8515: M=0.7\,M$_\odot$, f=0.1, 85\% carbon, 15\% silicates, 2.005\,$\mu$m$<$a$<$2.25\,$\mu$m, and M=0.5\,M$_\odot$, f=0.1, 85\% carbon, 3.005\,$\mu$m$<$a$<$3.25\,$\mu$m.}
\label{day8515figure}
\end{figure}

The effect on the predicted SED of increasing the dust size is shown in Figure~\ref{dustsizeeffect}.  We find that increasing the grain size can result in cooler dust as required by the observations, and a match to the observed SED can then be obtained by varying the dust mass and luminosity.  As the grain size is increased, the mass of dust required decreases, while the luminosity required increases.  For dust masses less than 1.0\,M$_\odot$, a grain size distribution in which grains larger than 2\,$\mu$m in radius are present is required.  Lower dust masses can also fit the observed SED but with even larger grain sizes required; to fit the data with a dust mass of 0.5\,M$_\odot$ requires grain radii of 3.005-3.25\,$\mu$m, while models with 0.4\,M$_\odot$ require grains larger than 5\,$\mu$m in radius.  0.4\,M$_\odot$ is the lowest possible dust mass which can still give a fit to the data; for lower dust masses, the luminosity required to match the {\it Herschel} observations is implausibly high, and also results in dust too hot, even for extremely large grain size distributions.  Dust masses higher than 0.7M$_\odot$ can give good fits to the SED with smaller grains; for a mass of 0.8M$_\odot$, micron-sized grains are required to fit the observed SED.  Figure~\ref{day8515figure} shows fits obtained with models containing 0.4-1.0\,M$_\odot$ of dust.

The best fitting day 615 model contained 4.15$\times$10$^{50}$ dust grains, while the best fitting models at day 8515 contain a much lower number: 0.4M$_\odot$ of dust with grain sizes between 5.005 and 5.25\,$\mu$m consists of 6.4$\times$10$^{41}$ grains, while 1.0M$_\odot$ of dust grains with 2.005$<$a$<$2.25\,$\mu$m consists of 2.26$\times$10$^{43}$ grains.  This suggests that the increase in grain size must be due not only to accretion, but also to coagulation of grains.

The large grains in these models have a narrow range of temperatures.  The distribution of grain temperatures throughout the model grids typically has a standard deviation of $\sim$ 1.5K.  In a day 9200 model (see Section~\ref{day9200section}) containing grains around 5\,$\mu$m in radius, consisting of 85\% amorphous carbon and 15\% silicates, with a mass of 0.8M$_\odot$, the dust temperatures were 15.7$\pm$1.4\, K and 19.2$\pm$1.5\,K for the carbon and silicates respectively.  The SED can thus be well fitted by a modified black body at a single temperature.  In comparison, in the best fitting model at day 615 (Section~\ref{day615section}), the temperature distributions were 252$\pm$29\,K for carbon, and 316$\pm$31\,K for silicates.

\begin{figure}
\includegraphics[width=0.47\textwidth]{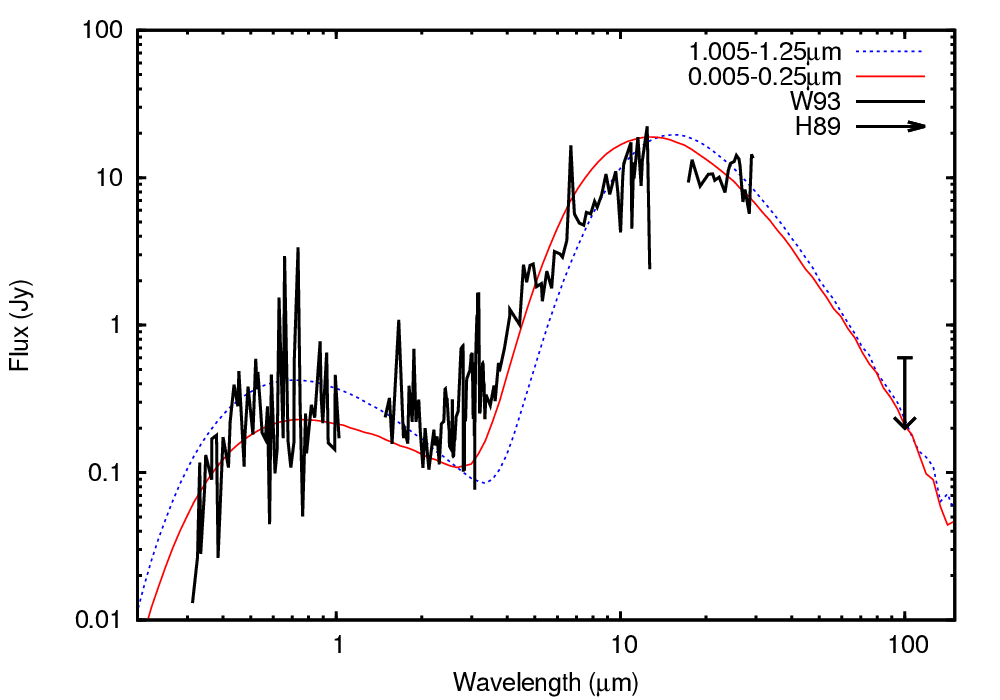}
\caption{Predicted SEDs at day 615 for a standard MRN dust size distribution with limits at 0.005 and 0.25\,$\mu$m, and for the enlarged grain size distribution required to fit later epochs.}
\label{nolargedust615}
\end{figure}

Having found that large dust grains are required for the dust in the models to be as cool as that observed at day 8515, we then investigated whether large grains could have been present at earlier epochs.  Rerunning our best fitting models at earlier times with the enlarged grain sizes required to fit later epochs, we find that the models no longer fit the data.  Figure~\ref{nolargedust615} shows that at day 615, using the enlarged grain size distribution required by late-time data results in less optical extinction and cooler dust, resulting in over-predicted optical fluxes and under-predicted near-IR fluxes.  The situation could be remedied by including more clumps in the model; however, one would then have to assume that the number of dusty clumps in the ejecta was not constant with time, and that more dusty clumps were present at earlier times than at later times.  It seems unlikely that dust could form in some clumps but then be destroyed, while in other clumps dust formation could be prolific.

The largest mass of dust that could still match the day 1153 observations was 0.003\,M$_{\odot}$, while the minimum amount of dust that must be present by day 8515 is 0.4M$_\odot$.  This means that more than 99\% of the dust in the ejecta formed after day 1153.  This observational result stands in potentially sharp contrast to recent theoretical work by \citet{2013ApJ...776..107S}, who predict that the bulk of dust formation should be complete within five years of the explosion.  For the two results to be reconciled, dust formation in the ejecta would have had to have occurred almost entirely in the narrow time window between 3.2 and 5 years after the explosion.  

\subsubsection{Dust composition}

Although there is observational evidence for iron and silicaceous dust formation in the remnant of SN\,1987A, our models show that any non-carbon dust which has formed by day 1153 must be less than 25\% of the total mass.  For the {\it Herschel} wavelengths observed on day 8515, including iron and/or silicate dust causes the dust to become slightly warmer, so that a higher dust mass is required to restore the fit to the SED.  A mass greater than 0.7\,M$_\odot$ may be difficult to reconcile with nucleosynthesis predictions for the explosion (\citealt{1990ApJ...349..222T,1988PASAu...7..355W}).  The dust is therefore likely to still be predominantly carbon at the latest epochs.  The models shown in Figure~\ref{day8515figure} used dust consisting of 85\% amorphous carbon and 15\% silicates, implying a total carbon mass of 0.42\,M$_\odot$ and a silicate mass of 0.075\,M$_\odot$ for the 0.5\,M$_\odot$ model.

We considered the effect on the derived mass of different optical constants for amorphous carbon; all masses quoted thus far were calculated from models in which the optical constants of \citet{1988ioch.rept...22H} were used.  To see if our results were sensitive to the choice of optical constants we re-ran all models using the optical constants from \citet{1996MNRAS.282.1321Z}, specifically the ACH2 sample produced by arc discharge between amorphous carbon electrodes in an H$_2$ atmosphere.  We find that the difference at early epochs is significant, with a fit to the observed SED being possible with a lower dust mass.  The Zubko optical constants have lower optical extinction such that the dust is cooler than found with the Hanner optical constants, and a smaller R$_{in}$ and R$_{out}$ are required to fit the SED.  At day 615, a fit to the observations can be obtained with M=5$\times$10$^{-4}$\,M$_\odot$ and R$_{in}$ and R$_{out}$ both 0.7$\times$ the value previously derived values.  At late times, however, the fit to the SED is less sensitive to the choice of optical constants.  For the parameters obtained using Hanner optical constants, re-running the model with Zubko optical constants does not change the predicted SED significantly.  For the models with smaller R$_{in}$ and R$_{out}$ found necessary to fit the day 615 observations using Zubko optical constants, when expanded to day 8515 we find that these models still require dust masses of the order of 0.6$\pm$0.2\,M$_\odot$ to fit the observed SED, and it remains the case that to obtain low enough dust temperatures while respecting the constraints on geometry obtained from the day 615 models, grain growth is required.  However, unless the luminosity is considerably lower than found using Hanner optical constants, the dust is too warm to fit the observations.  Our best fitting Zubko model had a luminosity of 315L$_\odot$, a dust mass of 0.65\,M$_\odot$, and grain sizes between 3.005 and 3.25\,$\mu$m.

\begin{figure*}
\includegraphics[width=0.47\textwidth]{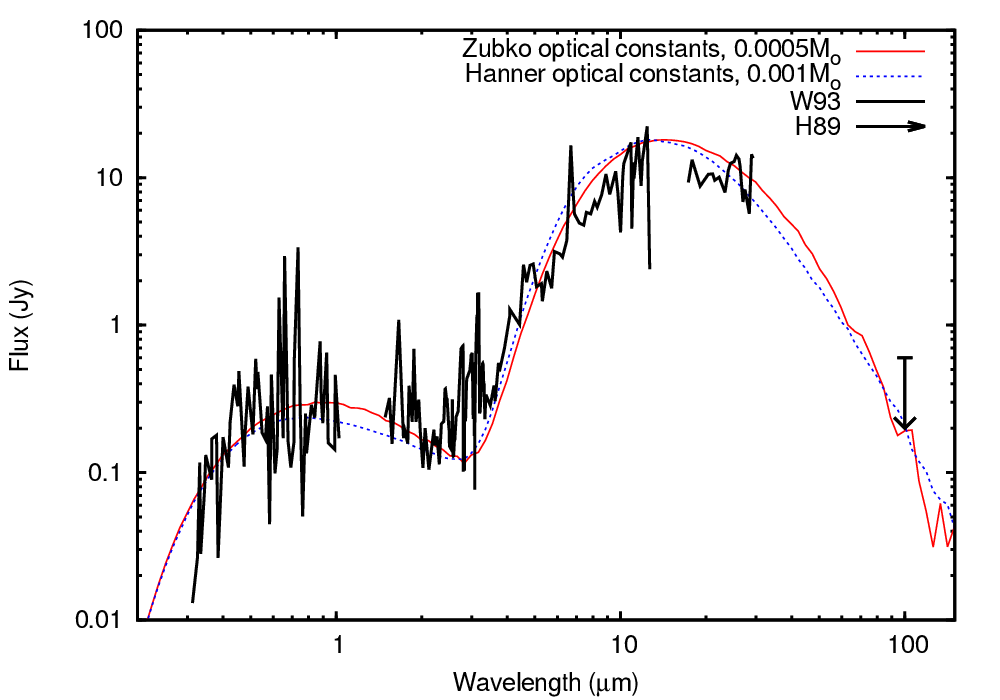}
\includegraphics[width=0.47\textwidth]{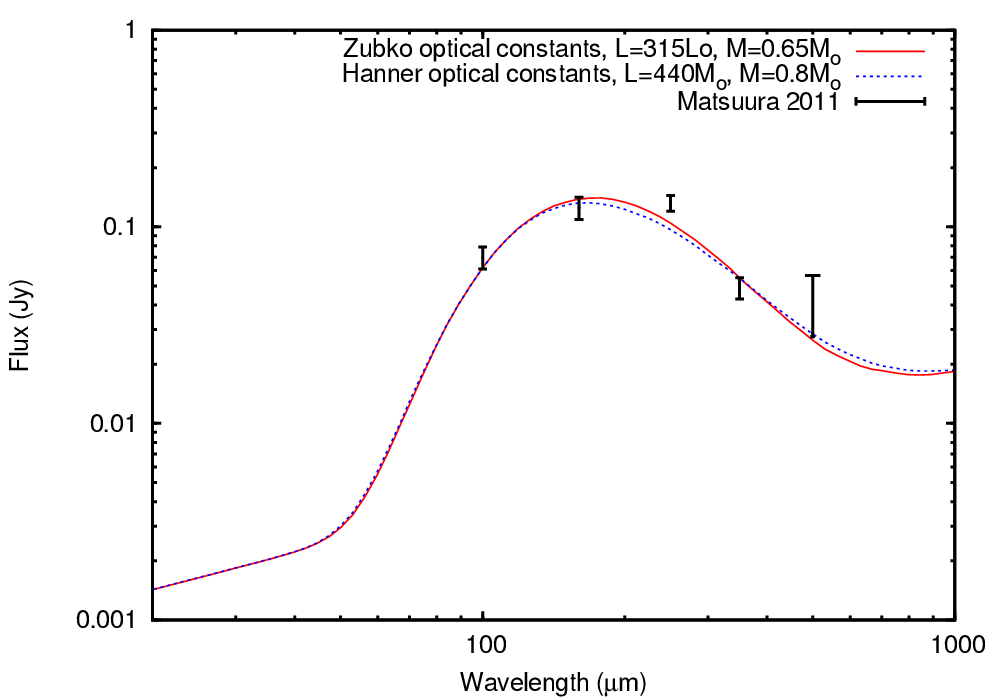}
\caption{Comparisons of models at days 615 (left) and 8515 (right) using different optical constants for amorphous carbon.  At day 615, models using Zubko optical constants require half the dust mass found using Hanner optical constants, and inner and outer radii reduced by $\times$0.7.  At day 8515, the mass required is similar for both sets of optical constants, but for the Zubko optical constants a lower luminosity is necessary to get cool enough dust to fit the observed SED.}
\label{opticalconstants}
\end{figure*}

Thus, while the amount of dust present at each epoch is fairly sensitive to the choice of optical constants, our findings that grain growth has occurred within the ejecta, and that the vast majority of the dust formed more than 1000 days after the explosion, are not affected by the choice of optical constants.  Figure ~\ref{opticalconstants} shows the predicted SEDs for both sets of optical constants, at days 615 and 8515 after the explosion.

\begin{table*}
\begin{tabular}{lllll}
\hline
Epoch (days) & R$_{in}$ ($\times$10$^{15}$\,cm) & R$_{out}$ ($\times$10$^{15}$\,cm) & Grain size distribution (a$_{min}$ ($\mu$m), a$_{max}$, ($\mu$m), power) & Estimated dust mass (M$_\odot$) \\
\hline
615  & 5.0 & 25.0 & 0.005-0.25, -3.5 & 0.0010 $\pm$ 0.0002 \\
775  & 6.3 & 31.5 & 0.005-0.25, -3.5 & 0.0020 $\pm$ 0.0002 \\
1153 & 9.4 & 46.9 & 0.005-0.25, -3.5 & 0.003 $\pm$ 0.001 \\
1300 & 10.6 & 52.9 & 0.005-0.25, -3.5 & 0.02--0.5 \\
8515 & 69.2 & 346.1 & 3.005-3.25, -3.5$^*$ & 0.6 $\pm$ 0.2 \\
9200 & 74.8 & 374.0 & 3.005-3.25, -3.5$^*$ & 0.8 $\pm$ 0.2 \\
\hline
\end{tabular}
\caption{The best fitting model parameters at each epoch.\newline$^*$At the final two epochs, the dust grain size distribution was created by increasing all grain radii by a constant amount, which means that the distribution of grain radii is proportional to (a-3.0)$^p$ rather than $a^p$.}
\label{dustmasstable}
\end{table*}

\subsection{Day 9200}
\label{day9200section}

We constructed a model at day 9200 after the explosion, to match {\it Herschel} photometry obtained by Matsuura et al. (submitted) on days 9090-9122 and ALMA 440 and 870\,$\mu$m photometry from days 9273-9387 (\citealt{2014ApJ...782L...2I}).  We estimated the synchrotron flux at submm wavelengths as before, by extrapolating the observations of \citet{2010ApJ...710.1515Z} to 9200 days, and then deriving a power law from those fluxes.  We find that the synchrotron flux in Jy can be approximated by the expression F$_\nu$($\lambda)$=0.411$\times (\frac{\lambda}{10^5})^{0.600}$\,Jy; we add the predicted synchrotron flux from this expression to our model fluxes for dust emission to obtain the total flux to be compared to the observations.  Once again, we restricted the volume filling factor of the models to the value of 0.1 required at early epochs, and we obtained a fit to observed fluxes by varying the dust mass, luminosity, and grain size distribution.

\begin{figure}
\includegraphics[width=0.47\textwidth]{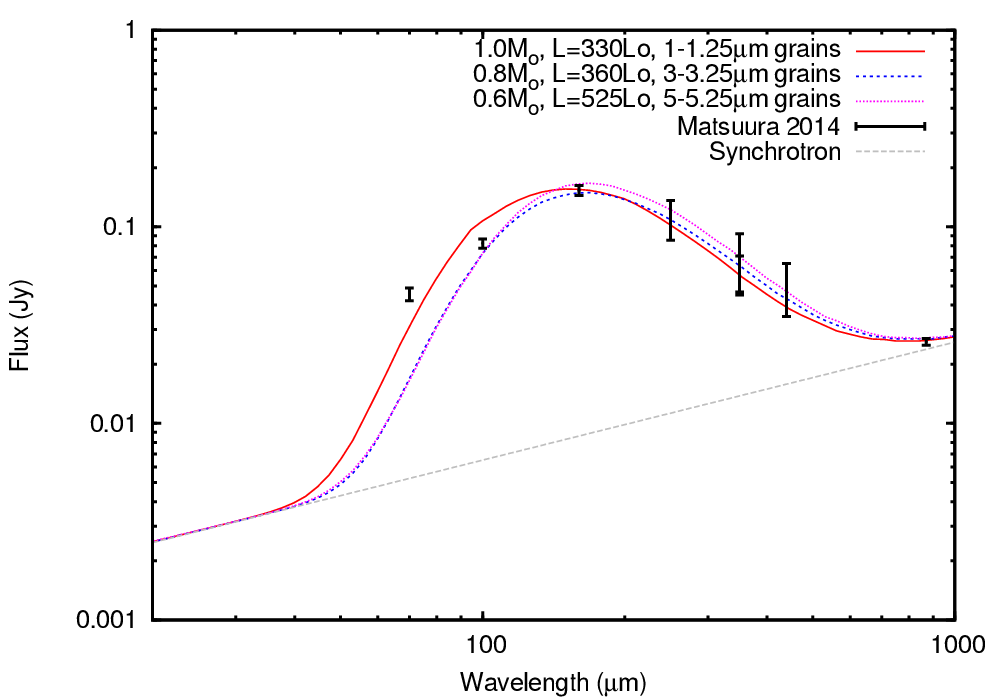}
\caption{Best fitting models for day 9200, with dust masses between 0.6 and 1.0M$_\odot$, and luminosities between 330 and 525\,L$_\odot$.  The dust is 85\% amorphous carbon and 15\% silicates.}
\label{day9200figure}
\end{figure}

We find that a slightly larger dust mass is required to fit the day 9200 observations than was required at day 8515.  Models with dust masses between 0.6 and 1.0M$_\odot$ can fit the observed data when the luminosity and grain size distribution are varied, and as before, the lower the dust mass, the higher the luminosity and the larger the dust grains required to fit the SED.  A good fit to the observations can be obtained with grains $\sim$3\,$\mu$m in radius, a luminosity of 360\,L$_\odot$ and a mass of 0.8\,M$_\odot$; fits can be obtained with dust masses up to 0.2\,M$_\odot$ above and below this value by varying the luminosity and the grain size distribution.  However, for dust masses as low as 0.6\,M$_\odot$, the luminosity required is 525L$_\odot$, which is higher than the luminosities required for any model on day 8200 and is thus probably unrealistic.  Figure~\ref{day9200figure} shows the predicted SEDs for three models.

\subsubsection{Clump size}

\begin{figure}
\includegraphics[width=0.47\textwidth]{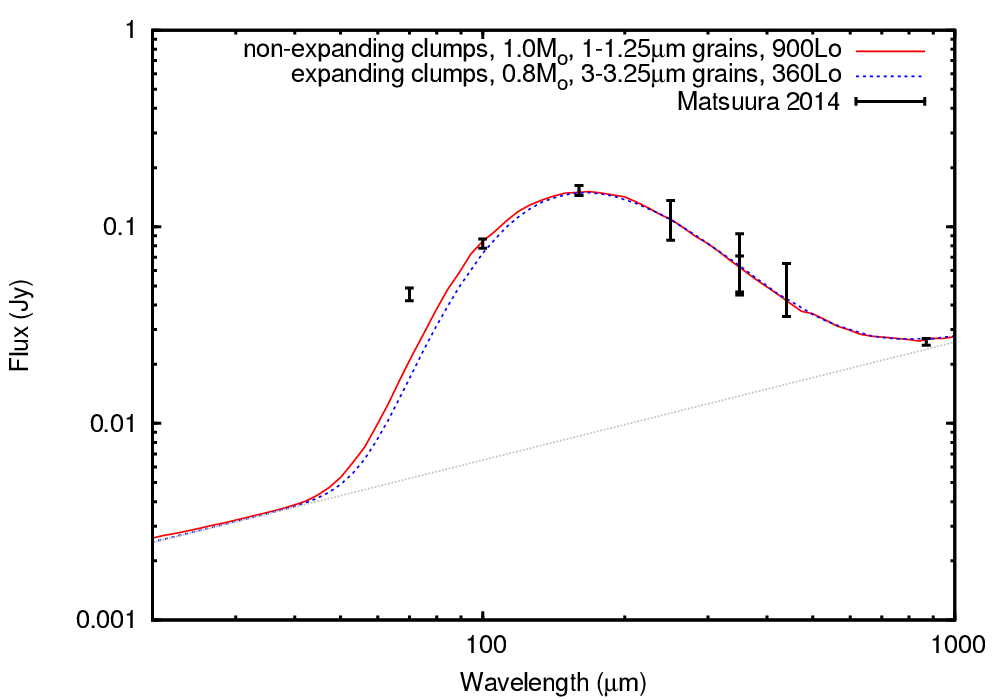}
\caption{Comparison of models at day 9200 with clump radius either held constant or allowed to expand uniformly.}
\label{clumpsizefig}
\end{figure}

For all the models discussed so far, all parts of the ejecta expanded uniformly between epochs.  We also considered models in which the clump radius is held constant, while the clumps themselves move outwards with the rest of the ejecta.  This may be a more realistic scenario, as in the case of uniform expansion the outer edge of the clumps would be moving at some 140\,km\,s$^{-1}$ relative to the centre of the clump, and after 23 years of expansion the density might be too low for continuing dust formation.  For the models with non-expanding clumps, a match to the observed SED requires a higher dust mass and a higher luminosity compared to the models in which the ejecta expands uniformly.  The greater density of the clumps results in cooler dust, and while an increase in grain radii is still necessary to fit the SED, the increase required is less than the case of uniformly expanding ejecta.  Figure~\ref{clumpsizefig} shows the predicted SEDs for two models at day 9200: for expanding clumps, the model contains 0.8\,M$_\odot$ of dust, with grain sizes from 3.005--3.25$\mu$m, and the luminosity is 360\,L$_\odot$, while in the case of non-expanding clumps, a similar fit to the SED is obtained with 1.0\,M$_\odot$ of dust, with grain sizes from 1.005--1.25$\mu$m, and a luminosity of 900\,L$_\odot$.  Such a high luminosity is difficult to account for: it would require that in addition to the 400\,L$_{\odot}$ from the decay of $^{44}$Ti, all of the $\sim$500\,L$_{\odot}$ X-ray luminosity from ejecta-CSM interaction goes into heating the dust.  This suggests that the clumps must be expanding but at a lower velocity than in the case of uniform expansion.  The dust masses derived assuming uniform expansion are thus likely to be somewhat underestimated.

\section{Discussion}

By constructing realistic three-dimensional models of the dusty ejecta of SN 1987A, we have provided new constraints on the location and timing of dust formation in the supernova ejecta.  We find that the very large dust mass deduced by \citet{2011Sci...333.1258M} from the far-infrared and submm emission detected by Herschel is physically plausible, and that the dust geometry determined from day 615 observations, expanded at constant velocity and with its total dust mass increased, then also gives good fits to the subsequent epochs.  At days 615, 775 and 1153, the observational data is sufficient to place strong upper limits on the mass of dust formed, and we thus show that the overwhelming majority of the dust mass present by day 8515 must have formed after day 1153.  The dust masses we derive at each epoch are summarised in Table~\ref{dustmasstable}.

Very few supernovae have been observed at the mid-infrared and longer wavelengths necessary to quantify dust production at epochs later than 1000 days after explosion.  In addition, at such late epochs, optical and infrared fluxes are commonly elevated by scattering and thermal echoes of the supernova flash from pre-existing circumstellar or interstellar material, which may dominate the observed fluxes at late times.  Otsuka et al. (2012) presented optical to mid-IR photometry of 6 supernovae at epochs $>$500 days after explosion, and found evidence of light echoes in 5 of the sample supernovae.  Given that we have shown that in the case of SN\,1987A, at least 99\% of the dust must have formed after day 1153, the much lower masses of dust found at mid-IR wavelengths for most core collapse supernovae observed within the first 1000 days clearly cannot rule out the later formation of a much greater mass of dust.

\citet{2013ApJ...776..107S} predicted that most dust formation in the ejecta of core collapse supernovae must occur at relatively early times ($<$5 years after the explosion), while for SN\,1987A we have shown that the vast majority of the dust must have formed more than 3.2 years after the explosion.  When the current large quantity of dust formed, and over how long, is not well constrained beyond the fact that it must have been after day 1153.  However, we find that the evolution of the dust mass with time can be approximated by a sigmoid function of the form

\begin{equation}
M_d(t) = ae^{be^{ct}}
\end{equation}

\begin{figure}
\includegraphics[width=0.47\textwidth]{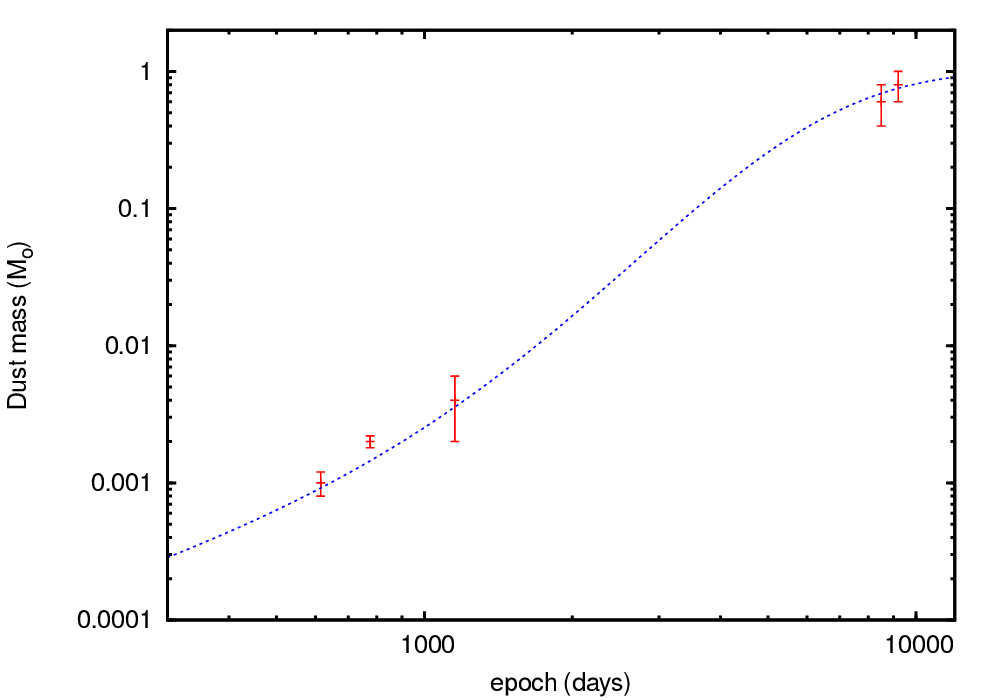}
\caption{Dust mass in the remnant of SN\,1987A v. time, with a sigmoid function fit overplotted.}
\label{dustfunction}
\end{figure}

From a least squares minimisation, we obtained coefficients of a=1.0\,M$_\odot$ (representing the maximum dust mass), b=-8.53 and c=-0.000366 to fit the dust masses estimated from our modelling.  The dust masses derived at each epoch, together with the sigmoid function, are shown in Figure~\ref{dustfunction}.  Such a function may be physically reasonable, with a slow increase in dust mass at first due to marginal fulfilment of the necessary physical conditions for dust formation, and a slow increase in the later stages of dust formation due to depletion of refractory elements, with the highest rate of dust formation occurring at an intermediate epoch.  If the dust formation can realistically be approximated by a function such as this, then even by 3000 days after the explosion, the dust mass was only 5\% of its final value, and 90\% of the dust would form between days $\sim$3000 and $\sim$14000.

We estimated the dust emission expected 7 and 10 years after the explosion by expanding the grid as for the other epochs, and using the dust mass estimated from the sigmoid function.  2500 days after the explosion, the expected dust mass is 0.03\,M$_\odot$, while 3650 days after the explosion, the function predicts a dust mass of 0.11\,M$_\odot$.  We ran models for luminosities of 750 and 1500\,L$_\odot$, and for three grain size distributions, in which grain radii were increased by 0, 0.05 and 0.5\,$\mu$m from the originally assumed grain size distribution.  We find that at day 2500, the SED peaks at around 100\,$\mu$m with fluxes ranging from 0.05--0.2\,Jy, while at day 3650, the predicted peak fluxes are 0.2--0.45\,Jy.

We have found that to obtain dust cool enough to match the observed SED at day 8515, large grains are required, with radii of up to several microns.  Considering the standard grain growth equation for the time taken for a grain to grow by accretion to a radius of $a$ from an initial radius $a_0$:

\begin{equation}
t = 4\rho_g \frac{(a-a_0)}{(n_i v_i A m_H S_i)},
\end{equation}

then for a time of 21.5 years between days 615 and 8515, a grain bulk density $\rho_g$ of 3.3\,gcm$^{-3}$, a sticking coefficient of 0.5, particles of atomic weight 12, and a gas temperature of 20K, the mean velocity of accreting particles $v_i$ is 2.3$\times$10$^4$cm\,s$^{-1}$, and the relation between the number density of accreting particles $n_i$ and the growth in grain radius $a - a_0$ in microns is given by:

\begin{equation}
n_i = 3.6\times10^{6}(a-a_0)
\end{equation}

For grains to grow in radius by 1\,$\mu$m in 21 years, the carbon atom number density required would be $\sim$3.6$\times$10$^{6}$\,cm$^{-3}$.  This would go up or down by a factor of ten for a final grain size an order of magnitude larger or smaller respectively.  At day 615, the clumps in our model have a radius of 8.3$\times$10$^{14}$cm and a volume of 2.4$\times$10$^{45}$cm$^3$; in the models with the best-fitting volume filling factor of 0.1 there are 12368 such clumps.  Allowing for other lighter species in the metal-rich regions of the ejecta, the total gas density would be $\sim$(1-3)$\times$10$^{7}$\,cm$^{-3}$.  An order of magnitude estimate for the total gas mass required in the clumps would then be 22.4\,M$_\odot$.  This exceeds the progenitor mass of 19\,M$_\odot$, again suggesting that the large grains present in the ejecta at very late times cannot have formed solely by accretion onto existing grains but must also have formed by coagulation.

Larger grains are more likely to survive the passage of the reverse shock through the ejecta; \citet{2007ApJ...666..955N} find that grains larger than 0.2\,$\mu$m in size would survive into the ISM essentially unchanged.  In our best fitting day 8515 and day 9200 models, all the grains are larger than this, and the majority of the dust could thus potentially survive the reverse shock.  A significant fraction of the dust can survive even in the models with smaller maximum radii, which fit the SED less well; for a model with grain radii $a$ in the range 0.05 $< $a$ < $ 0.5\,$\mu$m and n(a)$\propto$a$^{-3.5}$, for example, 2.8\% of the grains are larger than 0.2\,$\mu$m, but they contain 9.6\% of the mass.  This suggests, therefore, that a significant fraction of the dust formed in the ejecta of SN\,1987A will survive the passage of the reverse shock.

Recently, Gall et al. (2014) have found evidence for large grains ($>$0.7\,$\mu$m) being formed in SN2010jl.  They find evidence for large grains within a short time of the supernova explosion, in contrast to our finding for SN1987A that the larger grains present at day 8500 could not have been present before day 1000.  Nevertheless, in both cases, the presence of large dust grains that will probably survive the passage of the reverse shock through the ejecta supports the hypothesis that core-collapse supernova can be a major source of dust in galaxies.

One of the long standing questions regarding SN\,1987A is the nature of the compact remnant.  Despite extensive searches, no evidence for a pulsar has been found at any wavelength, despite strong theoretical predictions that one should have formed in the explosion (\citealt{1993ARA&A..31..175M,1989ARA&A..27..629A}).  \citet{2005ApJ...629..944G} found an upper limit to the integrated flux between 0.29--0.965\,$\mu$m of F$<$1.6$\times$10$^{-14}$ erg\,s$^{-1}$cm$^{-2}$, and used this limit to rule out scenarios including accretion onto a neutron star or black hole, based on an assumption of relatively low dust absorption in the remnant, with A$_v$=0.595.  The large amount of dust now known to be present means that the optical extinction could be much higher than previously assumed.  If one or more dense clumps of dust were to lie along the line of sight, the upper limit on the optical flux of the compact remnant could be much higher than the value found by \citet{2005ApJ...629..944G}.  In a day 8515 model with 0.6M$_\odot$, each clump has a mass of 4.85$\times$10$^{-5}$M$_\odot$, and an edge-to-edge optical depth of 12.2, equivalent to A$_V$=13.3; assuming that the dust clumps expand with $t$, the optical depth will decline with 1/$t^2$.  If a single clump were to lie directly along the line of sight, then with no dust formation or destruction, or alteration of the dust geometry apart from expansion, the lower extinction assumed by \citet{2005ApJ...629..944G} would not be reached for approximately another 85 years.  In fact, several clumps are likely to lie along the line of sight; in the geometry adopted, the clump covering factor is almost unity.  From the point of view of the central source, the average number of clumps intercepted by an arbitrary line of sight is 3.9, and the chance of a line of sight not intercepting any clumps is 0.1\%.  We find that over all lines of sight, the median value of A$_v$ is 4.8, and it exceeds 15 in approximately 5\% of directions.  The vast dust reservoir of SN\,1987A may thus keep any compact remnant hidden from view at optical wavelengths for a very long time.

\section{Acknowledgments}

We thank Paul Harvey and Allan Meyer for providing further information about the day 1179 and 1183 {\em KAO} 50-$\mu$m observations of SN~1987A.  This work was co-funded under the Marie Curie Actions of the European Commission (FP7-COFUND).

\bibliographystyle{mn2e}
\bibliography{sn1987a}

\end{document}